# Tuning perpendicular magnetic anisotropy in the MgO/CoFeB/Ta thin films

T. Zhu*, P. Chen, Q. H. Zhang, R. C. Yu, and B. G. Liu

Institute of Physics and Beijing National Laboratory for Condensed Matter Physics, Chinese Academy of Sciences, Beijing 100190, P.R. China

*Correspondence and requests for materials should be addressed to T.Z. (tzhu@aphy.iphy.ac.cn)

**Abstract**

**Understanding the magnetic anisotropy at ferromagnetic metal/oxide interface is a fundamental and intriguing subject. Here we propose a new approach to manipulate the strength of perpendicular magnetic anisotropy (PMA) by varying MgO thickness in the MgO/CoFeB/Ta thin films. We identify that the PMA at the MgO/CoFeB interface can be tuned by the interface structure. We find that the strength of PMA decreases dramatically with the increasing of MgO thickness due to the onset of crystalline MgO forming. Furthermore, we demonstrate a large linear anomalous Hall effect in the annealed MgO/CoFeB/Ta thin film with thick MgO layer. Our work opens a new avenue to manipulate the magnetic anisotropy by the modification of the ferromagnetic metal/oxide interface.**

# 1. Introduction

The characteristic of magnetic anisotropy is one of the keys to determine the applications of the ferromagnetic thin films for decades.[1-15] Due to the interfacial magnetic anisotropy the direction of magnetization changes from in-plane to out-of-plane and induces a perpendicular magnetic anisotropy (PMA) in the ferromagnetic thin film. The PMA has been found in the conventional ferromagnetic metal/ non-magnetic metal (FM/NM) multilayers, such as Co/Pd multilayers,[5] and in the annealed ferromagnetic metal/oxide thin films, such as Pt/Co/AlO$_x$[6] and Ta/CoFeB/MgO[7] thin films. For the later film, recent theoretical calculations of PMA at Fe(Co)/MgO suggested that the origin of PMA is attributed to the overlap between O-2p and transition metal 3d orbitals.[8,9] Experimentally, the strength of PMA can be tuned not only by varying the thickness of FM layer, but also by changing the oxidation status at the interface of FM/MO$_x$,[10-13] or even by applying an electric field.[7,14] Since the magnetic anisotropy ultimately arises from spin-orbit interaction, it is of great interest in both experimental and theoretical views to investigate the mechanism of PMA at the MgO/CoFeB interface, which have received continuous attentions due to their potential use in highly sensitive magnetic field sensors or in spin-transfer-torque magnetic random access memories.[15-18]

In this paper, we propose a new approach to manipulate the strength of PMA at the MgO/CoFeB interface. It is found that the strength of PMA dramatically decreases with the increasing of MgO thickness in the MgO/CoFeB/Ta thin films, due to the onset of crystalline MgO forming, which opens a new avenue to manipulate the magnetic anisotropy of the ferromagnetic metal/oxide interface.

## 2. Experimental process

MgO (1-5)/CoFeB (0.8-3)/Ta (0.55-5) samples were deposited on a thermally oxidized silicon wafer by magnetron sputtering with a base pressure of $5\times10^{-6}$ Pa. The numbers in brackets are nominal thicknesses in nanometers. Ar was used as the sputtering gas and the sputtering pressure for metals was 0.4 Pa. MgO layers was deposited by radiofrequency (RF) sputtering and the sputtering pressure was 0.2 Pa. High purity of Ta (99.95%), MgO (99.99%), and $Co_{40}Fe_{40}B_{20}$ (99.9%, Functional Materials International, Japan) were used as the target materials. All the samples were capped with a 3-nm-thick MgO to prevent oxidation. The microstructure was characterized by using high resolution transmission electron microscopy (HRTEM) in a FEI Tecnai F20. The cross-sectional specimen for TEM observation was prepared by $Ar^+$ ion-beam milling with low-energy, low-angel parameter and liquid-nitrogen cooled stage at Gatan Model 691 PIPS after mechanical grinding and subsequent dimpling. Here, the PMA properties have been investigated by using anomalous Hall effect (AHE) which is widely used to characterize perpendicular ferromagnetic films.[6-11,19,20] The Hall resistivity measurements were measured by using standard cross-structure four points in a physical property measurement system (PPMS, Quantum Design PPMS-14T) within the temperature range of 5 - 300 K. The longitudinal resistivity was measured in the same sample from another channel of PPMS at same time. The magnetoresistance in all films at 5 T is smaller than 0.3%. Room temperature AHE was also measured in another commercial Hall probing system by using an electromagnet (up to 1.5 T) and high resolution multimeters (Keithley® 220 and 2182). To well eliminate the ordinary Hall effect, $R_{AH}$ is defined as the zero field extrapolation of the high field data.

## 3. Results

The stack structures of MgO (1-5)/CoFeB (1.2)/Ta (1.1) were deposited on the thermally oxidized silicon substrates (Si/SiO$_2$) by magnetron sputtering. The numbers in brackets are nominal thicknesses in nanometers. The patterned Hall bar samples were annealed from 90 to 450 °C for 1 hour. Similar with the previous works,[15,16] obvious PMA has been observed when CoFeB thickness ranging from 1.04 to 1.46 nm annealed at 300 °C. In this work, we chose $t_{CoFeB}$ = 1.2 nm to further study the tunable PMA in the MgO/CoFeB/Ta thin films.

Figure 1 show the sample structure of the AHE measurement and the representative AHE data for an MgO/CoFeB/Ta sample with 1.5-nm-thick MgO annealed at 300 °C. For a perpendicular FM film, the magnetic easy-axis is perpendicular to the film plane. AHE measures the transverse voltage with respect to the transport direction depends on the spontaneous magnetization along the perpendicular direction.[21] Thus, the normalized out-of-plane magnetization ($M_Z$) is equal to the normalized anomalous Hall resistance ($R_{xy}$). The PMA properties of a perpendicular FM thin film can be identified by the $R_{xy}$-$H$ hysteresis loop with an external magnetic field perpendicular to the film. As shown in Fig. 1(b), the $R_{xy}$-$H$ loop is square, indicating a pronounced large PMA with 100 % out-of-plane remanence.

It is well known that the effective magnetic anisotropy energy density ($K_{eff}$) is used to characterize the strength of magnetic anisotropy of a magnetic thin film. The $K_{eff}$ in the first order approximate can be defined as

$$K_{eff} = K - 2\pi M_S^2, \tag{1}$$

where $K$ is magnetic anisotropy energy density including volume anisotropy and interfacial anisotropy, and $M_S$ the saturation magnetization. A perpendicular easy axis of magnetization appears when $K_{\text{eff}} > 0$. It should be pointed that the value of $K_{\text{eff}}$ includes higher order contributions of uniaxial anisotropy. Experimentally, $K_{\text{eff}}$ calculates from the enclosed area between the perpendicular and in-plane magnetization curves,[22]

$$K_{\text{eff}} = \frac{H_K M_S}{2}, \tag{2}$$

where $H_K$ is anisotropy field measured along the hard axis. $H_{K\perp}$ is the perpendicular anisotropy field for a perpendicular FM thin film, which can be obtained by in-plane anomalous Hall measurement with an in-plane external field,[23,24] measuring the decrease of $M_Z$ due to the in-plane external field, and calculating the in-plane magnetization as $|M_x| = \sqrt{1 - M_z^2}$. As shown in Fig. 1, $H_{K\perp} = 2.4$ kOe for the annealed MgO/CoFeB/Ta with 1.5-nm-thick MgO layer. Thus, we can use the $H_{K\perp}$ to characterize the PMA properties of the perpendicular MgO/CoFeB/Ta thin film,[16] assuming that the volume anisotropy is neglected.[15]

Figure 2 shows the representative AHE data for the MgO/CoFeB/Ta sample with 4-nm-thick MgO annealed at 300 $^\circ$C. The perpendicular coercivity of the film is quite low (about 20 Oe), which exhibiting a much weak PMA. Similarly, $H_{K\perp}$ was also probed by AHE applying an in-plane external magnetic field. It is surprising that $H_{K\perp}$ is only 0.66 kOe for the MgO/CoFeB/Ta thin films with 4-nm-thick MgO layer, which reflecting that the PMA strength of the MgO/CoFeB/Ta thin films strongly depends on the MgO thickness.

Figure 3 shows the phase diagram of $H_{K\perp}$ as a function of annealing temperature ($T_{\text{ann}}$) and MgO thickness ($t_{\text{MgO}}$) for the MgO ($t_{\text{MgO}}$)/CoFeB (1.2)/Ta (1.1) samples. Apparently, the maximum of $H_{K\perp}$ appears when annealing temperature is around 300$^\circ$C and MgO thickness around 1.5 nm. To simplify the analysis, we use a zero of $H_{K\perp}$ platform ($H_{K\perp} = 0$) to present the area with $H_{K\perp} \leq 0$ and then focus on the detail of the film with out-of-plane magnetization remanence. From Eq. 1, $K_{eff}$ is the result of the competition between the demagnetization energy and the perpendicular magnetic anisotropy energy. As for the annealed MgO/CoFeB/Ta thin films, the magnitudes of $M$s usually increase with the increasing of $T_{ann}$, due to the crystallization of CoFeB triggered by a decrease of boron concentration during annealing.[20] This is to say that the demagnetization term ($2\pi M_S^2$) will increase during annealing. Thus, the fact of the increase of $H_{K\perp}$ indicates that the contribution of CoFeB/MgO interface dominates the PMA properties with the increasing of $T_{ann}$, which consistent with the previous works.[15-18,20,25] On the other hand, the degradation of PMA can be seen when $T_{ann}$ is above 330°C due to the interdiffusion of Ta,[26] which results in such a non-monotonic PMA behavior.

Figure 4a shows the representative $H_{K\perp}$ and $K_{eff}$ as a function of $t_{MgO}$ for the MgO ($t_{MgO}$)/CoFeB (1.2)/Ta (1.1) annealed at 300 °C, which clearly exhibiting the exist of an optimized MgO thickness, *i. e.* $t_{MgO} = 1.5$ nm, for large PMA properties of MgO/CoFeB/Ta thin films. It is clear that $H_{K\perp}$ depends not only on $T_{ann}$, but also on $t_{MgO}$. $H_{K\perp}$ decreases dramatically when $t_{MgO}$ is thicker than 1.5 nm. To further investigate the PMA mechanism arising from the CoFe(B)/MgO interface, the films' microstructure was characterized by using high resolution transmission electron microscopy (HRTEM). Figure 4b shows the representative HRTEM images. The 1.5-nm-thick MgO in the MgO/CoFeB/Ta thin film is amorphous, whereas 4-nm-thick MgO is partially polycrystalline. This feature reveals that the microstructure of sputtered MgO layer strongly depends on the MgO thickness in the annealed MgO/CoFeB/Ta thin film, which will be discussed later.

The $M_S$ may change with the increasing of MgO thickness due to the onset of crystalline MgO forming. The change of magnetization in the CoFeB is mainly due to the presence of a MDL at the interface during deposition or intermixing upon annealing.[27,28] However, we did not observe an obvious change in the magnetization with the increase of MgO thickness at the same annealing temperature, within the accuracy of magnetization measurement. To take into account the absolute magnetization $M_S$ = 663 emu/cm$^3$ for the MgO/CoFeB/Ta thin films,[20] $K_{eff}$ is calculated according to the Eq. 2 as shown in Fig. 4a. The highest $K_{eff}$ is 0.78×10$^6$ erg/cm$^3$ for the MgO/CoFeB/Ta thin film with 1.5-nm-thick MgO layer. However, the $K_{eff}$ is only 0.22×10$^6$ erg/cm$^3$ for the MgO/CoFeB/Ta thin film with 4-nm-thick MgO layer, which is only 28% of the highest $K_{eff}$. Hence, the $K_{eff}$ dramatically changes with the varying of MgO thickness, which indicating the MgO thickness is also a key parameter to tune the PMA properties of the MgO/CoFeB/Ta thin films.

On the other hand, a linear $R_{xy}$-$H$ loop can be obtained when $K_{eff}$ < 0. As an example of potential application, we demonstrate a large linear anomalous Hall effect in MgO/CoFeB/Ta thin films by simply varying the MgO thickness. The Hall resistance in a FM thin film consists two terms: one presents the ordinary Hall effect and another presents the AHE which is much larger than the former one. As shown in Figure 5, the linear Hall field sensitivity ($S_{AHE}$) is defined as the ratio of $R_{AH}$ to $H_S$, $R_{AH}/H_S$. Here, $R_{AH}$ is defined as the zero field extrapolation of the high field anomalous Hall resistance in order to well eliminate the ordinary Hall effect and $H_S$ the out-of-plane saturation field. Due to large $H_S$ (> 10 kOe) of a typical magnetic thin film, its $S_{AHE}$ is no more than several Ω/kOe.[29] In Fig. 5, the representative $R_{xy}$ - $H$ loop of MgO (4)/CoFeB (1.2)/Ta (1.1) thin film annealed at 150 $^o$C. The $S_{AHE}$ at room temperature is 1900 Ω/kOe due to the small $H_S$ of the sample.

## 4. Discussion

There is no direct first principle calculation to study the PMA mechanism of the MgO/CoFeB/Ta thin films, because the CoFeB is usually amorphous or partial polycrystalline. However, one can still qualitatively investigate the origin of PMA in MgO/CoFeB/Ta system based on the first principle calculation of Fe|MgO(100) or CoFe|MgO(100).[8,9,31,32] Both the experimental and theoretical works indicate that the interface of CoFeB/MgO plays a key role in the PMA of the MgO/CoFeB/Ta thin films.

Our results show that the crystalline degree of MgO layer in the MgO/ CoFeB/Ta thin film depends on the MgO thickness. The sputtered very thin MgO layer (~ 1 nm) may keep its amorphous status. Indeed, it has been reported that the crystalline degree of the sputtered MgO layer depends on the substrates or buffer layers. For example, it has been reported that the thin sputtered MgO layer on Si(100) was amorphous.[33] On the other hand, the thick sputtered MgO layer in the MgO/CoFeB/Ta thin film exhibits partial crystalline (Fig. 4b). Accompanied with the onset of the MgO crystallization with the increasing of MgO thickness, the PMA significantly reduces. The tunable $K_{eff}$ turns to a negative value but very close to 0. And then, a large linear Hall field sensitivity is observed.

Such a PMA mechanism can qualitatively explain the recent experimental result on the reduction of PMA for the CoFeB directly deposited on MgO(100) single crystal substrate,[34] although the model needs more experimental proof in future. Further calculating PMA in CoFe(B)/MgO(100) needs to set up a suitable model, for example a model based on an ideal CoFe bcc structure with boron substituting for a lattice site and the possible interface configuration.[35] Nevertheless, the PMA can be tuned by the local configuration at interface, which is affected by the crystalline degree of MgO layer.

## 5. Conclusion

We report a new mechanism to manipulate the perpendicular magnetic anisotropy at the MgO/CoFeB interface. It is found that the strength of PMA decreases dramatically in the annealed MgO/CoFeB/Ta thin film due to the onset of the crystalline MgO forming, which can be a potential candidate for highly sensitive magnetic field sensor, memory, or logic device.

**ACKNOWLEDGEMENTS**

This work has been supported by the National Basic Research Program of China (2012CB933102) and National Science Foundation of China (Grants 11079052, 11174354, and 11174359).

**Figure 1**

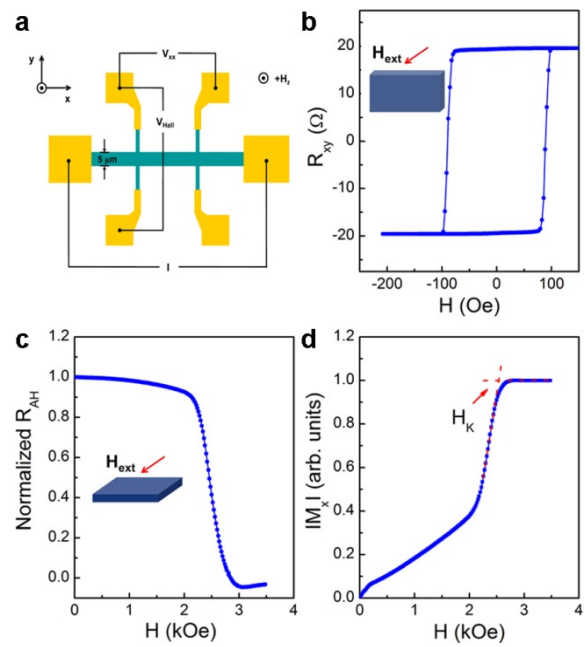

**Figure 1 | Measurement geometry and AHE data. a**, Sample structure of the AHE measurements. **b**, The representative AHE loop for the MgO/CoFeB/Ta sample with 1.5-nm-thick MgO annealed at 300$^o$C, which exhibits a pronounced large PMA. **c**, The in-plane AHE curve. **d**, The calculated normalized |$M_x$| as a function of external magnetic field, in which the perpendicular anisotropy field ($H_{K\perp}$) is obtained.

**Figure 2**

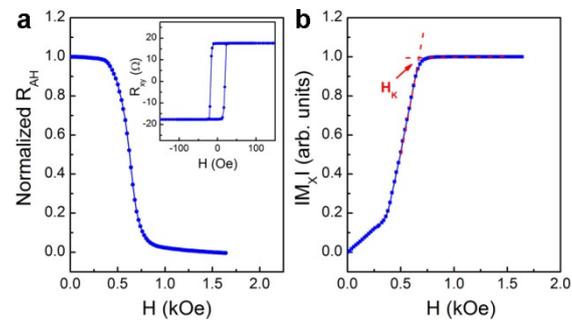

**Figure 2 | The AHE data for the film with thick MgO thickness. a**, The representative in-plane AHE loop for the MgO/CoFeB/Ta thin film with 4-nm-thick MgO annealed at 300 °C, exhibiting a weak PMA. The inset shows the AHE loop. **b**, The calculated normalized |$M_x$| as a function of external magnetic field.

**Figure 3**

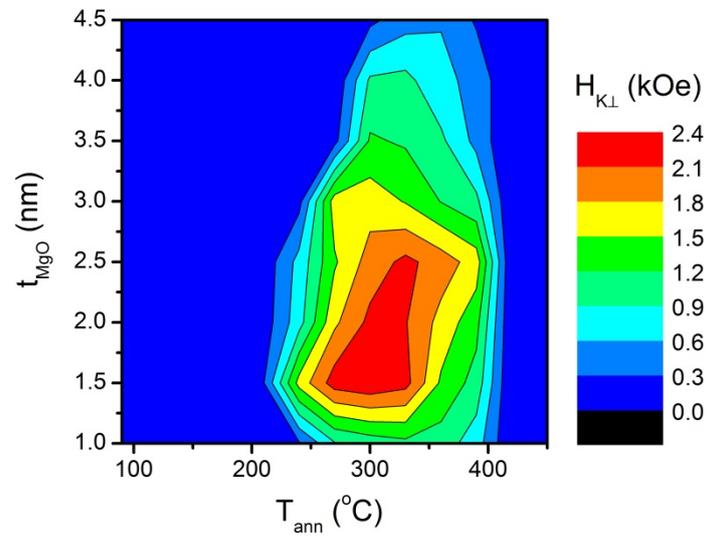

**Figure 3 | The phase diagram of the perpendicular anisotropy field.** The perpendicular anisotropy field as a function of annealing temperature and MgO thickness for MgO ($t_{MgO}$)/CoFeB (1.2)/Ta (1.1) thin films.

**Figure 4**

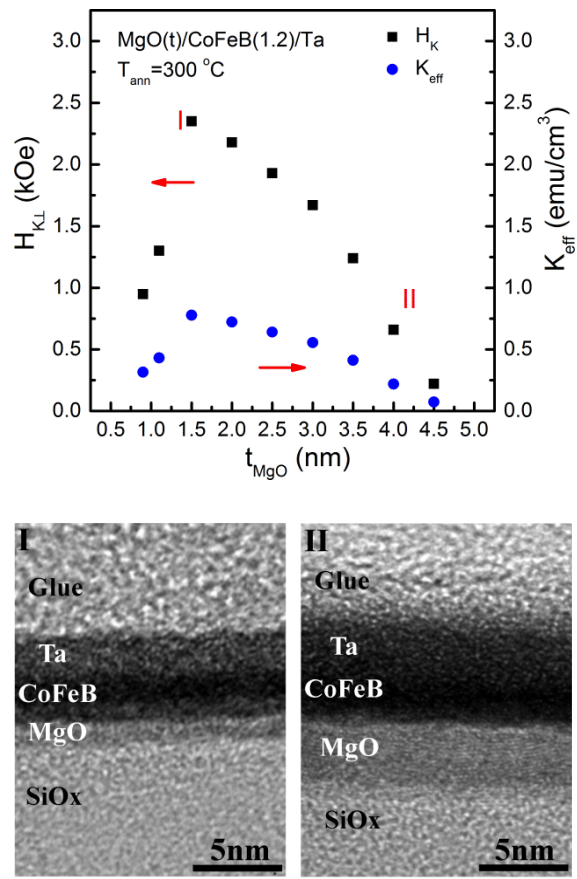

**Figure 4 | The MgO thickness dependence of perpendicular anisotropy field and microstructure. a**, The perpendicular anisotropy field as a function of MgO thickness for MgO ($t_{MgO}$)/CoFeB (1.2)/Ta (1.1) thin films annealed at 300 °C. **b**, The representative HRTEM images for the MgO/CoFeB/Ta thin films with $t_{MgO}$ = 1.5 (I) and 4 nm (II). 1.5-nm-thick MgO shows amorphous, whereas 4-nm-thick MgO shows polycrystalline.

**Figure 5**

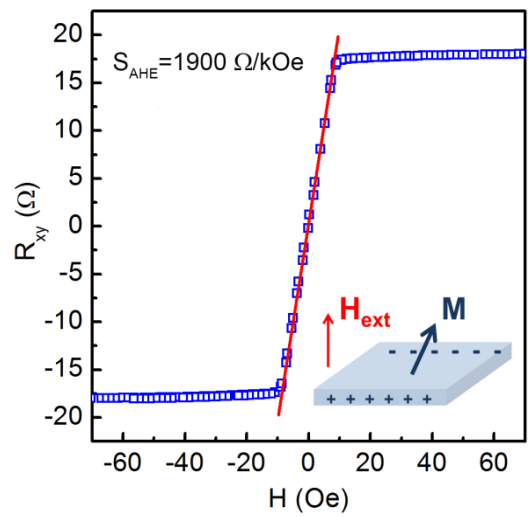

**Figure 5 | The representative curves of $R_{xy}$ vs $H$ for MgO (4)/CoFeB (1.2)/Ta (1.1) film ($T_a$ = 150°C).** The $S_{AHE}$ = 1900 Ω/kOe is measured.